\documentclass[preprint,5p,sort&compress]{elsarticle}

\usepackage{graphicx,epsfig,amssymb,times}
\usepackage{amsmath,amsfonts,amssymb}
\usepackage{bm}
\usepackage{bbold}
\usepackage{epstopdf}
\usepackage[linktocpage,colorlinks]{hyperref}
\usepackage[caption=false]{subfig}
\usepackage[usenames]{color}
\usepackage{graphicx,epsfig,amssymb} %include figure files
\usepackage{amsmath,amsfonts, times}
\usepackage{bm} %include bold math: \bm{} creates bold letters in math mode

\usepackage{epstopdf}
\usepackage[linktocpage,colorlinks]{hyperref}
\usepackage[caption=false]{subfig}
\usepackage[usenames]{color}
\usepackage{natbib}
\usepackage{soul}
\usepackage[utf8x]{inputenc}

\definecolor{coolblack}{rgb}{0.0, 0.18, 0.39}
\definecolor{darkred}{rgb}{0.5,0,0}
\definecolor{darkgreen}{rgb}{0,0.5,0}
\definecolor{darkblue}{rgb}{0,0,0.5}
\definecolor{lapislazuli}{rgb}{0.15, 0.38, 0.61}
\definecolor{venetianred}{rgb}{0.78, 0.03, 0.08}
\definecolor{bleudefrance}{rgb}{0.19, 0.55, 0.91}
\definecolor{dogwoodrose}{rgb}{0.84, 0.09, 0.41}
\hypersetup{colorlinks=true, citecolor=darkgreen, linkcolor=darkblue,
	urlcolor = blue}

\def\be{\begin{equation}}
\def\ee{\end{equation}}

\newcommand{\bea}{\begin{eqnarray}}
\newcommand{\eea}{\end{eqnarray}}
\newcommand{\ben}{\begin{enumerate}}
	\newcommand{\een}{\end{enumerate}}
\newcommand{\bi}{\begin{itemize}}
	\newcommand{\ei}{\end{itemize}}

\newcommand{\bqa}{\begin{eqnarray}}
\newcommand{\eqa}{\end{eqnarray}}

\def\ga{\mathrel{\raise.3ex\hbox{$>$\kern-.75em\lower1ex\hbox{$\sim$}}}}
\def\la{\mathrel{\raise.3ex\hbox{$<$\kern-.75em\lower1ex\hbox{$\sim$}}}}

\def\be{\begin{equation}}
\def\ee{\end{equation}}

\def\I_M{{I_{\scriptscriptstyle M\times M}}}

\def\be{\begin{equation}}
\def\ee{\end{equation}}
\def\bea{\begin{eqnarray}}
\def\eea{\end{eqnarray}}
\newcommand{\beq}{\begin{eqnarray}}
\newcommand{\eeq}{\end{eqnarray}}

\usepackage{graphicx,epsfig,amssymb} %include figure files
\usepackage{amsmath,amsfonts, times}
\usepackage{bm} %include bold math: \bm{} creates bold letters in math mode

\usepackage{epstopdf}
\usepackage[linktocpage,colorlinks]{hyperref}
\usepackage[caption=false]{subfig}
\usepackage[usenames]{color}
\usepackage{natbib}
\usepackage{soul}
\usepackage[utf8x]{inputenc}

\definecolor{coolblack}{rgb}{0.0, 0.18, 0.39}
\definecolor{darkred}{rgb}{0.5,0,0}
\definecolor{darkgreen}{rgb}{0,0.5,0}
\definecolor{darkblue}{rgb}{0,0,0.5}
\definecolor{lapislazuli}{rgb}{0.15, 0.38, 0.61}
\definecolor{venetianred}{rgb}{0.78, 0.03, 0.08}
\definecolor{bleudefrance}{rgb}{0.19, 0.55, 0.91}
\definecolor{dogwoodrose}{rgb}{0.84, 0.09, 0.41}
\hypersetup{colorlinks=true, citecolor=darkgreen, linkcolor=darkblue,
	urlcolor = blue}

\def\be{\begin{equation}}
\def\ee{\end{equation}}

\def\be{\begin{equation}}
\def\ee{\end{equation}}

\def\be{\begin{equation}}
\def\ee{\end{equation}}
\def\bea{\begin{eqnarray}}
\def\eea{\end{eqnarray}}

\begin{document}
\begin{frontmatter}
\title{\boldmath The Constraint of ${H_0}$ from Galaxy clusters and Hubble parameter data}

\author[a]{Hai Huang\corref{mycorrespondingauthor} }
\cortext[mycorrespondingauthor]{Corresponding author}
\ead{ lapulandebaby@126.com}
\author[b]{Long Huang }

\address[a]{College of Mechanical Engineering, Guiyang University, Guiyang, Guizhou 550005, People’s Republic of China}
\address[b]{ College of Physics, Hunan University of Science and Technology, Xiangtan 411100, Hunan Province, China.}

\begin{abstract}
Using comoving distance $d_c$ and angular diameter distance $d_A$, we recalculate parameters describing kinematical state of the universe, still combining the kinematical model of universe but not relying on dynamical equations for gravity. Comoving distance $d_c$ comes from Hubble data H(z) and is more reliable. Angular diameter distance $d_A$ comes from SZE (Sunyaev-Zel dovich Effect) and X-ray data, and needs calibration. In low redshift case, we use expansion of relation between luminosity distance and redshift about redshift $z$; in high redshift case, we take variable substitution $y=1/(1+z)$, and expand the relation between luminosity distance and redshift about variable $y$ in order to reduce computational errors. Finally we get the more precise value of Hubble parameter $H_0=69.13\pm 0.24{\kern 1pt} km{\kern 1pt}  \cdot {s^{ - 1}} \cdot Mp{c^{ - 1}}$, corresponding to 0.4\% uncertainty in $68.3\%$ confidence region, also deceleration factor $q_0=-0.57\pm0.04 $ and acceleration rate $j_0=1.28\pm0.33$, and their statistical values and probability graph. We compare the values of ${H_0}$, ${q_0}$, ${j_0}$ with those obtained from other observation data and model.
\end{abstract}

\begin{keyword}
Hubble parameter\sep Hubble constant\sep Galaxy clusters
\end{keyword}

\end{frontmatter}

\section{Introduction}

 The value of Hubble parameter $H_0$, deceleration factor $q_0$, acceleration rate $j_0$ are very importance to cosmology, it make us learn about the kinematics state of the universe. At present, the cosmological computational methods used to constrain the cosmological parameter are mainly two kinds, one is using dynamical equations for gravity and combining the observational data, getting the Hubble parameter $H_0$, the current matter density $\rho_{m_0}$, dark energy density $\rho_\Lambda$, and calculating to obtain the deceleration factor $q_0$ by the current matter density and dark energy density; the other is independently using the kinematical model of universe and not relying on dynamical equations for gravity. On the data side, the calibration of data is crucial for the parameters constraint result. Using five Cepheid-based distances to local Type Ia SNe host galaxies, supplemented with an indirectly inferred distance to NGC 3627, Saha et al.\citep{1} derived zero points to uncorrected B- and V -band Hubble diagrams. Ignoring external systematic effects, their resulting Hubble constant was $H_0=58\pm3{\kern 1pt} km{\kern 1pt}  \cdot {s^{ - 1}} \cdot Mp{c^{ - 1}}$. Using only the uncorrected Calan-Tololo sample, but including the I-band data, and ignoring those calibrators with poor photometry or without a direct Cepheidbased distance determination, both Hamuy et al. and Riess et al.\citep{2} found $H_0=56\pm2{\kern 1pt} km{\kern 1pt}  \cdot {s^{ - 1}} \cdot Mp{c^{ - 1}}$, in agreement with Saha et al.\citep{1}. Now more and more cosmological observation data including Cepheid\citep{3,4} , TRGB\citep{5}, SBF\citep{6}, Tully-Fisher Relation\citep{7,8} are used to calibrate SNe Ia observation data\citep{9}. Mould and Sakai calibrated Tully-Fisher Relation using 14 TRGB data, and fitting $H_0=73\pm 5 {\kern 1pt} km{\kern 1pt}  \cdot {s^{ - 1}} \cdot Mp{c^{ - 1}}$(only statistic error), 7\% uncertainty\citep{8}; Riess et al used Maser Galaxy NGC4258 data, and combined with eight Cepheid data and SNe Ia data in the same Galaxy , mading calibration to Cepheid data and SNe Ia data, then combined with the cosmography to get the Hubble parameter $H_0 = 73.8 \pm 2.4 {\kern 1pt} km{\kern 1pt}  \cdot {s^{ - 1}} \cdot Mp{c^{ - 1}}$ including systematic error, corresponding to 3.3\% uncertained\citep {10}, Blakeslee et al. also used SBF data and got $H_0 = 73\pm4 \pm11 {\kern 1pt} km{\kern 1pt}  \cdot {s^{ - 1}} \cdot Mp{c^{ - 1}}$, corresponding to 5.5\% and 15\% uncertained\citep {11}. On the other hand, geometric distance observation data including Maser Galaxy data\citep{12,13}, angular diameter distance $ d_A $ from SZE and X-ray data\citep{14,15}, comoving distance $ d_c $ from $H(z)$ data\citep{16} are used to constrained cosmological parameters. Bonamente et al. used $d_A$ data from SZE and X-ray observation data, and combined with $\Lambda CDM $model to get $H_0=76.9_{ - 3.4{\kern 1pt}  - 8.0}^{ + 3.9{\kern 1pt}  + 10}{\kern 1pt} km{\kern 1pt}  \cdot {s^{ - 1}} \cdot Mp{c^{ - 1}}$,  respectively corresponding to 5\% and 11\% uncertainty in statistic and systematic error\citep {14}. Jimenez et al. used $H (z) $ data, and combined with $\omega(z) CDM $ model to fit to get $H_0 = 69 \pm 12 {\kern 1pt} km{\kern 1pt}  \cdot {s^{ - 1}} \cdot Mp{c^{ - 1}}$,  20\% uncertainty\citep {17}. Meanwhile Plank Collabration recently used CMB data publishing by Plank probes, and combined multipole power spectrum coefficient scalar formula with $\Lambda CDM$ model to get the Hubble parameter $H(z)$\citep{18,19}.

 Our work is mainly to use $d_A$ data from SZE and X-ray data and $d_c$ data from $H(z)$ data, and combining with cosmography to fit to get the more precise value of the Hubble parameter $H_0$, deceleration factor $q_0$, and acceleration rate $j_0$.  The Hubble parameter $H (z) $ data is mainly obtained by measuring different red envelope galaxies age and redshift, and  combining with the relation of Hubble parameter $H(z)$ and redshift $z $\citep {17} ; At the same time, H(z) data can be obtained from the BAO scale as a standard ruler in the radial direction known as "Peak" Method \citep {20}. The measurement of clusters angle diameter distance $d_A$ provide a convenience for us to konw the cluster geometrical distance, $d_A$ data is mainly obtained from the X-ray and SZE data. Bonamente et al. took The Hydrostatic Equilibrium model to get 38 $d_A$ data by the X-ray and SZE data, and Filippis et al. used the Spherical -$\beta$ model to get 25 $d_A$ data\citep{14,15}

In the second section, we have analyzed $H (z) $ data , and getting the comving distance $d_c$ by the relation of ${d_c} $and $H (z) $ \citep{26}. At the same time, we introduce two group cluster angle diameter distance $d_A$ which based on two different models of Elliptical -$\beta$ model and Spherical -$\beta$ model  \citep {14,15}. In the third section, we expand the relation of luminosity distance $d_l$ and redshift $z $ about the variable $z $ and the variable $y = 1/1+z$, and getting two type different expansion form of the relation of the luminosity distance and redshift, meanwhile we introduce the relation of $d_l $ and $d_A$, $d_c$. In the fourth section, we take MCMC technology, combining $d_c $, $d_A$ observations with two different expansion form of luminosity distance-redshift relation to fit and get the Hubble constant $H_0$, decelerate factor $q_0$, accelerate rate $j_0$ best fitting values, statistic average values and probability distribution. In section 5, we make a comparison for the parameter values from the different data and the model and analyze to get the conclusion. In section 6, we discuss the paper.

\section{The observation data analysis}

$H (z) $ data is mainly obtained by measuring different red-envelope galaxy age and redshift, at the same time it also can be obtained according to BAO scale as a standard ruler in the radial direction known as "Peak" Method . Now we get 28 in the redshift of $0.07 \le z \le 1.75$ $H (z) $data, where Simon et al. used the relation of $H(z)$ and redshift $z$, time $t$ and got nine H(z) data on the basis of Jimenez et al. \citep {21}, the Hubble parameter $H(z)$ can be written in the form of
\begin{equation}\label{eq1}
H(z) =  - \frac{1}{{1 + z}}\frac{{dz}}{{dt}}
\end{equation}
Meanwhile Stern et al. \citep{16} revised these data at 11 redshifts from the differential ages of red-envelope galaxies. Gaztanaga et al. \citep{20} took the BAO scale as a standard ruler in the radial direction, obtained two data. Recently, Moresco et al. \citep{22} obtained 8 data from the differential spectroscopic evolution of early-type galaxies as a function of redshift. Blake et al. \citep{23} obtained 3 data through combining measurements of the baryon acoustic peak and Alcock-Paczynski distortion from galaxy clustering in the WiggleZ Dark Energy Survey. Zhang et al. \citep{24} obtained another 4 data.

The relation of comoving distance $d_c$ and Hubble parameters $H(z)$ can be written in the form of\citep{25}
\begin{equation}\label{eq2}
d\_c = c\int_0^z {\frac{{dz'}}{{H(z')}} \approx \frac{c}{2}\sum\limits_{i = 1}^n {({z_{i + 1}} - {z_i})\left[ {\frac{1}{{H({z_{i + 1}})}} + \frac{1}{{H({z_i})}}} \right]} }
\end{equation}

Because the value of $H_0$ have a very small influence on the calculation of $d_c$, even more less than the error of $d_c$, so it can not affect the fitting values of parameters. We assumes $H_0=73.8{\kern 1pt} km{\kern 1pt}  \cdot {s^{ - 1}} \cdot Mp{c^{ - 1}}$\citep{15}, and get 28 $d_c$ data. Meanwhile, we combine with comving distance error $\sigma_{d_c}$ formule
\begin{equation}\label{eq3}
\sigma _{{d_c}}^2 = \frac{c}{2}\sum\limits_{i = 1}^N {({z_{i + 1}} - {z_i})} {(\frac{{\sigma _{{H_{i + 1}}}^2}}{{H_{i + 1}^4}} + \frac{{\sigma _{{H_i}}^2}}{{H_i^4}})^{1/2}}
\end{equation}
and getting comving distance error $\sigma_{d_c}$.

\section{The expansion of the luminosity distance and redshift relation }
\subsection{Cosmography \uppercase\expandafter{\romannumeral1}}

According to cosmological principle, Robertson-Walker metric is taken as \citep{26}
\begin{equation}\label{eq4}
d{s^2} =  - d{t^2} + {a^2}(t)\left. {\left\{ {\frac{{d{r^2}}}{{1 - k{r^2}}}} \right. + {r^2}d{\theta ^2} + {r^2}{{\sin }^2}{\kern 1pt} {\kern 1pt} \theta d{\phi ^2}} \right\},
\end{equation}
where $a(t)$ is the cosmic scale factor, $k$ is the curvature term. A star which is located at $r_1$, $\theta$, and $\phi$, at moment $t_1$, emits a beam of light to Earth, and it reaches at moment $t_0$, namely
\begin{equation}\label{eq5}
c\int_{{t_1}}^{{t_0}} {\frac{{dt}}{{a(t)}}}  = \int_{{r_1}}^0 {\frac{{dr}}{{\sqrt {1 - k{r^2}} }}} {\kern 1pt} .
\end{equation}
Equation (5) yields
\begin{equation}\label{eq6}
r = \left\{ \begin{split}
&\sin \int_{{t_1}}^{{t_0}} {\frac{{dt}}{{a(t)}}{\kern 1pt} {\kern 1pt} {\kern 1pt} {\kern 1pt} {\kern 1pt} {\kern 1pt} {\kern 1pt} {\kern 1pt} {\kern 1pt} {\kern 1pt} {\kern 1pt} {\kern 1pt} {\kern 1pt} {\kern 1pt} {\kern 1pt} {\kern 1pt} {\kern 1pt} {\kern 1pt} {\kern 1pt} {\kern 1pt} {\kern 1pt} {\kern 1pt} {\kern 1pt} {\kern 1pt} {\kern 1pt} {\kern 1pt} k =  + 1,} \\
&\int_{{t_1}}^{{t_0}} {\frac{{dt}}{{a(t)}}{\kern 1pt} {\kern 1pt} {\kern 1pt} {\kern 1pt} {\kern 1pt} {\kern 1pt} {\kern 1pt} {\kern 1pt} {\kern 1pt} {\kern 1pt} {\kern 1pt} {\kern 1pt} {\kern 1pt} {\kern 1pt} {\kern 1pt} {\kern 1pt} {\kern 1pt} {\kern 1pt} {\kern 1pt} {\kern 1pt} {\kern 1pt} {\kern 1pt} {\kern 1pt} {\kern 1pt} {\kern 1pt} {\kern 1pt} {\kern 1pt} {\kern 1pt} {\kern 1pt} {\kern 1pt} {\kern 1pt} {\kern 1pt} {\kern 1pt} {\kern 1pt} {\kern 1pt} {\kern 1pt} {\kern 1pt} {\kern 1pt} {\kern 1pt} k = 0{\kern 1pt} {\kern 1pt} {\kern 1pt} {\kern 1pt} {\kern 1pt} {\kern 1pt} {\kern 1pt} {\kern 1pt} {\kern 1pt} {\kern 1pt} ,} \\
&\sinh \int_{{t_1}}^{{t_0}} {\frac{{dt}}{{a(t)}}{\kern 1pt} {\kern 1pt} {\kern 1pt} {\kern 1pt} {\kern 1pt} {\kern 1pt} {\kern 1pt} {\kern 1pt} {\kern 1pt} {\kern 1pt} {\kern 1pt} {\kern 1pt} {\kern 1pt} {\kern 1pt} {\kern 1pt} {\kern 1pt} {\kern 1pt} {\kern 1pt} {\kern 1pt} {\kern 1pt} k =  - 1.}
\end{split} \right.{\kern 1pt} {\kern 1pt} {\kern 1pt}
\end{equation}

According to the relation of scale factor $a(t)$  and redshift $z$ , namely
\begin{equation}\label{eq7}
1 + z = \frac{{a({t_0})}}{{a({t_1})}},
\end{equation}
the relation of the luminosity distance and redshift is\citep{26}
\begin{equation}\label{eq7}
{d_l} = a({t_0}){r_1}(1 + z){\kern 1pt} {\kern 1pt} ,
\end{equation}
Making Taylor expansion to scale factor $a(t)$, combine equations (3), (4), (5), and get the series expansion of relation of the luminosity distance and redshift \citep{27}
\begin{equation}\label{eq9}
\begin{split}
{d_l} &= \frac{c}{{{H_0}}}\{ z + \frac{{ - {q_0} + 1}}{2}{z^2} - \left[ {\frac{{ - 3{q_0}^2 - {q_0} + {j_0} + {\Omega _{{k_0}}} + 1}}{6}} \right]{z^3},\\
\end{split}
\end{equation}
where
\begin{equation}\label{eq10}
\begin{split}
&{H_0} = {\left. {\frac{{\dot a(t)}}{{a(t)}}} \right|_{t = {t_0}}}{\kern 1pt} {\kern 1pt} ,\\
&{q_0} = {\left. { - \frac{1}{{{H^2}}}\frac{{\ddot a(t)}}{{a(t)}}} \right|_{t = {t_0}}}{\kern 1pt} {\kern 1pt} ,\\
&{j_0} = {\left. {\frac{1}{{{H^3}}}\frac{{{a^{(3)}}(t)}}{{a(t)}}} \right|_{t = {t_0}}}{\kern 1pt} {\kern 1pt} ,\\
&{\Omega _{{k_0}}} = \frac{k}{{{H_0}^2{a^2}({t_0})}}{\kern 1pt} {\kern 1pt} {\kern 1pt} {\kern 1pt} .
\end{split}
\end{equation}

This is the expansion of luminosity distance and redshift about $z$ (Cosmography \uppercase\expandafter{\romannumeral1}).

\subsection{Cosmography \uppercase\expandafter{\romannumeral2}}

We use variable substitution $y=1/1+z$ and make series expansion of the relation (6) of the luminosity distance and redshift about variable $1-y$ .
When $t_0-t_1$ is small quantity, from equation(6), we make Taylor expansion to radial distance $r_1$ for the case $k=\pm1$ and get
\begin{equation}\label{eq11}
 \begin{split}
{r_1} &= \left[ {\int_{{t_1}}^{{t_0}} {\frac{{cdt}}{{a(t)}}} } \right] - \frac{k}{{3!}}{\left[ {\int_{{t_1}}^{{t_0}} {\frac{{cdt}}{{a(t)}}} } \right]^3} + \frac{{{k^2}}}{{5!}}{\left[ {\int_{{t_1}}^{{t_0}} {\frac{{cdt}}{{a(t)}}} } \right]^5}\\
{\kern 1pt} {\kern 1pt} {\kern 1pt} {\kern 1pt} {\kern 1pt} {\kern 1pt} {\kern 1pt} {\kern 1pt} {\kern 1pt}  &+ O\left( {{{\left[ {\int_{{t_1}}^{{t_0}} {\frac{{cdt}}{{a(t)}}} } \right]}^7}} \right).
\end{split}
\end{equation}
We also make Taylor expansion to the scale factor $a(t)$ and we get
\begin{equation}\label{eq12}
\begin{split}
a(t) &= a({t_0})\{ 1 + {H_0}(t - {t_0}) - \frac{{{H_0}^2{q_0}}}{2}{(t - {t_0})^2}\\
{\kern 1pt}  &+ \frac{{{H_0}^3{j_0}}}{{3!}}{(t - {t_0})^3} \} ,
\end{split}
\end{equation}
Substituting equation (12) into (11), we get
\begin{equation}\label{eq13}
\begin{split}
 - a({t_0}){r_1} &= ({t_1} - {t_0}) - \frac{1}{2}{H_0}{({t_1} - {t_0})^2}\\
{\kern 1pt} {\kern 1pt} {\kern 1pt} {\kern 1pt} {\kern 1pt} {\kern 1pt} {\kern 1pt} {\kern 1pt} {\kern 1pt} {\kern 1pt} {\kern 1pt} {\kern 1pt} {\kern 1pt} {\kern 1pt} {\kern 1pt} {\kern 1pt} {\kern 1pt} {\kern 1pt} {\kern 1pt}  &+ \left[ {\frac{1}{3}{H_0}^2(\frac{1}{2}{q_0} + 1) - \frac{k}{{3!{a^2}({t_0})}}} \right]{({t_1} - {t_0})^3}\\
{\kern 1pt} {\kern 1pt} {\kern 1pt} {\kern 1pt} {\kern 1pt} {\kern 1pt} {\kern 1pt} {\kern 1pt} {\kern 1pt} {\kern 1pt} {\kern 1pt} {\kern 1pt} {\kern 1pt} {\kern 1pt} {\kern 1pt} {\kern 1pt} {\kern 1pt} {\kern 1pt} {\kern 1pt} & + \left[ {\frac{1}{4}{H_0}^3( - \frac{{{j_0}}}{6} - {q_0} - 1) + \frac{k}{{4{a^2}({t_0})}}{H_0}} \right]{({t_1} - {t_0})^4}\\
\end{split}
\end{equation}
Let $y=1/1+z$ and it is known that the relation of the scale factor $a(t)$ and redshift is given as
\begin{equation}\label{eq14}
1 + z = \frac{{a({t_0})}}{{a({t_1})}}.
\end{equation}
From equations (12) and (14), we get
\begin{equation}\label{eq15}
\begin{split}
y &= \frac{{a({t_1})}}{{a({t_0})}} = 1 + {H_0}({t_1} - {t_0}) - \frac{{{H_0}^2{q_0}}}{2}{({t_1} - {t_0})^2}\\
{\kern 1pt} {\kern 1pt} {\kern 1pt} {\kern 1pt} {\kern 1pt} {\kern 1pt} {\kern 1pt} {\kern 1pt} {\kern 1pt} {\kern 1pt} {\kern 1pt} {\kern 1pt} {\kern 1pt} {\kern 1pt} {\kern 1pt} {\kern 1pt} {\kern 1pt} {\kern 1pt} {\kern 1pt} {\kern 1pt} {\kern 1pt} {\kern 1pt} {\kern 1pt} {\kern 1pt} {\kern 1pt} {\kern 1pt} {\kern 1pt} {\kern 1pt} {\kern 1pt} {\kern 1pt} {\kern 1pt} {\kern 1pt} {\kern 1pt} {\kern 1pt} {\kern 1pt} {\kern 1pt} {\kern 1pt} {\kern 1pt} {\kern 1pt} {\kern 1pt} & + \frac{{{H_0}^3{j_0}}}{{3!}}{({t_1} - {t_0})^3}
\end{split}
\end{equation}
Exchanging function $y-1$ and variable $t_1-t_0$ in the above power series, we get
\begin{equation}\label{eq16}
\begin{split}
{t_1} - {t_0} &= \frac{1}{{{H_0}}}(y - 1) + \frac{{{q_0}}}{{2{H_0}}}{(y - 1)^2} + \frac{{3{q_0}^2 - {j_0}}}{{3!{H_0}}}{(y - 1)^3}\\
\end{split}
\end{equation}
The relation of luminosity distance and redshift is given by
\begin{equation}\label{eq17}
{d_l} = a({t_0}){r_1}(1 + z){\kern 1pt} {\kern 1pt} .
\end{equation}
From equations (13), (16) and (17), we get the power series expansion of the luminosity distance $d_l$ about $1-y$ by the expression
\begin{equation}\label{eq18}
\begin{split}
{d_l} &= \frac{c}{{{H_0}y}}\{ (1 - y) - \frac{{{q_0} - 1}}{2}{(1 - y)^2} + [\frac{{3{q_0}^2 - 2{q_0} - {j_0}}}{6}\\
{\kern 1pt} {\kern 1pt} {\kern 1pt} {\kern 1pt} {\kern 1pt} {\kern 1pt} {\kern 1pt} {\kern 1pt} {\kern 1pt} {\kern 1pt} {\kern 1pt} {\kern 1pt}  &+ \frac{{ - {\Omega _{{k_0}}} + 2}}{6}]{(1 - y)^3}{\kern 1pt}
\end{split}
\end{equation}

This is the expansion of luminosity distance and redshift about $y$ (Cosmography \uppercase\expandafter{\romannumeral2})

\section{The fitting of parameters}

We now use $d_A$ data from Elliptical $\beta$ model and spherical $\beta$ model and $d_c$ data from the relation of $H(z)$ and $d_c$, meanwhile combining with Cosmography \uppercase\expandafter{\romannumeral1} and Cosmography \uppercase\expandafter{\romannumeral2} to fit the parameters. We take weighted mean method to 2 same redshift data when using Elliptical $\beta$ model $d_A$ data, finally, getting 24 in the redshift range of  $0.023 \le z \le 0.784$ data;

We don't consider the influence of cosmic opacity\citep{28,29}, and assuming DD relation is\citep{30,31}
\begin{equation}\label{eq19}
  {d_l} = {d_A}{(1 + z)^2}
\end{equation}
and
\begin{equation}\label{eq20}
  {d_l} = {d_c}(1 + z)
\end{equation}

We use $\chi^2$ statistic fitting method to constrain the parameters
\begin{equation}\label{eq21}
\begin{split}
\chi _{cluster + H(z)}^2 &= \sum\limits_{cluster} {\frac{{d_{l,obs}^{clusters}({z_i}) - {d_l}({H_0},{q_0},{j_0};{z_i})}}{{\sigma _{clsuter}^2({z_i})}}} \\
&+ \sum\limits_{H(z)} {\frac{{d_{l,obs}^{H(z)}({z_i}) - {d_l}({H_0},{q_0},{j_0};{z_i})}}{{\sigma _{H(z)}^2({z_i})}}}
\end{split}
\end{equation}

Where $d_{l,obs}^{cluster}(z)$ is given by equation(19), $d_{l,obs}^{H(z)}({z})$ is given by equation(20), and ${d_l}({H_0},{q_0},{j_0};{z})$ is given by equation(9, 18), error ${\sigma _{cluster}}(z) = {\sigma _{{d_A}}}(z){(1 + z)^2}$, ${\sigma _{H(z)}}(z) = {\sigma _{{d_c}}}(z)(1 + z)$.

We first combine $d_A$ and $H(z)$ data with Cosmography \uppercase\expandafter{\romannumeral1}, and constraining the parameters. Because the curvature team $\Omega _{{k_0}}$ is very small, we assume $\Omega _{{k_0}}=0$, and then using MCMC technology and $\chi^2$ statistic fitting method to get the parameters $H_0$, $q_0$, $j_0$ best fitting values, statistic mean and error in $68.3\%$($\Delta {\chi ^2} \le 1$) confidence region (see table 1), and probability density graph, confidence region of plane ($H_0$, $j_0$)(see figure 1, 2)

\begin{table*}
\footnotesize
 \centering
  \caption{The best fitting and mean values of the cosmological parameters from $d_A$, $d_c$ and $H(z)$observation data combined with the Cosmography \uppercase\expandafter{\romannumeral1}, in 68.3\% confidence region.}
  \vspace{0.3cm}
  \begin{tabular}{@{}cccccc@{}}
  \hline
  & Data parameters&$H_0$&$q_0$&$j_0$&$\chi _{\min }^2$\\ \hline
   Elliptical-$\beta$ model &Best fitting &74.43   &    -0.72      &    -0.7     &    24.19$/$24\\
                            &Mean&74.6$\pm$2.41 & -0.62$\pm$0.05   &   -0.18$\pm$2.14  \\ \hline
   Corrected Spherical &Best fitting &65.91  &-0.61    &    -0.02   &    31.48$/$32\\
                            -$\beta$ model&Mean&66.91$\pm$3.09 & -0.75$\pm$0.46   &   1.33$\pm$3.03  \\ \hline
  $H(z)$ &Best fitting &67.73   &    -0.29      &    -0.1    &    32.98$/$28\\
                            &Mean&67.75$\pm$0.22 & -0.29$\pm$0.02   &   -0.09$\pm$0.05  \\ \hline
   Elliptical-$\beta$ model &Best fitting &67.76   &    -0.29      &    -0.09     &    58.56$/$52\\
                           $+H(z)$  &Mean&67.8$\pm$0.21 & -0.29$\pm$0.02   &   -0.08$\pm$0.04  \\ \hline
   Corrected Spherical &Best fitting &67.66   &    -0.28      &    -0.11     &    68.31$/$60\\
                            -$\beta$ model $+H(z)$&Mean&67.77$\pm$0.2 & -0.29$\pm$0.02   &   -0.09$\pm$0.04  \\
  \hline
\end{tabular}
\end{table*}

\begin{figure*}[htbp]
\includegraphics[scale=0.7]{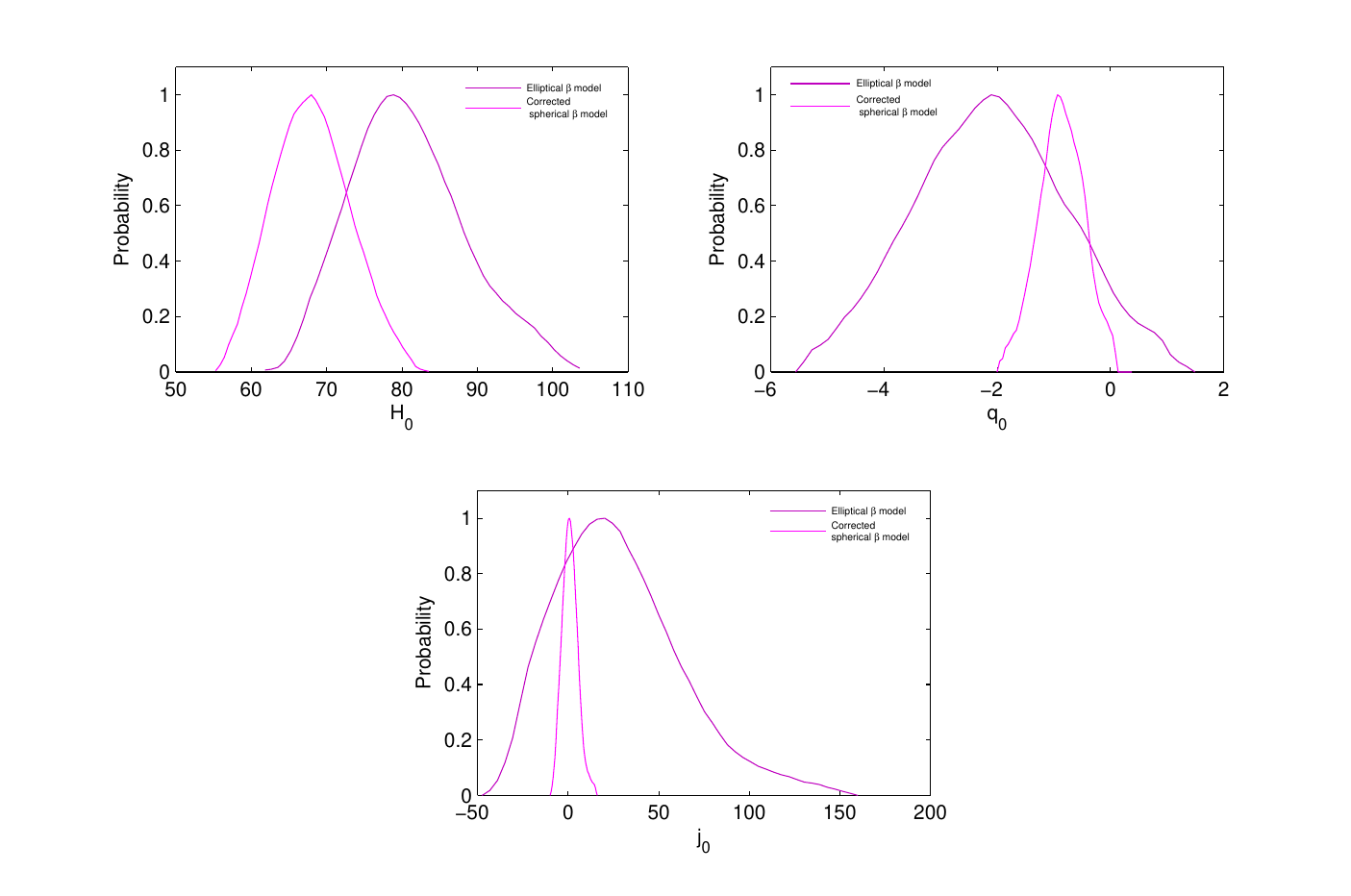}
  \caption{One-dimensional probability distribution graph for the parameter $H_0$, $q_0$ and $j_0$ from  $d_A$ observation data combined with the Cosmography \uppercase\expandafter{\romannumeral1}}.
\end{figure*}

\begin{figure*}[htbp]
\includegraphics[scale=0.7]{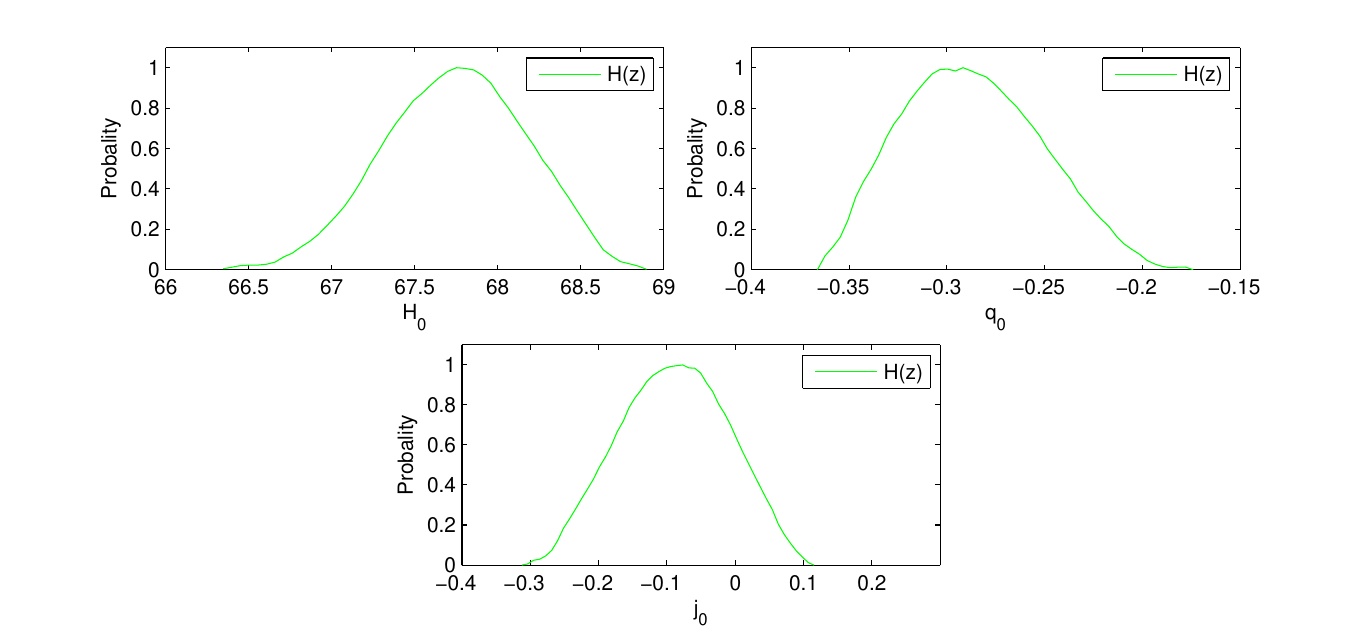}
  \caption{One-dimensional probability distribution graph for the parameter $H_0$, $q_0$ and $j_0$ from  $H(z)$,$d_A+H(z)$ observation data combined with the Cosmography \uppercase\expandafter{\romannumeral1}}.
\end{figure*}

\begin{figure}[tbp]
\centering
\includegraphics[angle=0,scale=0.4]{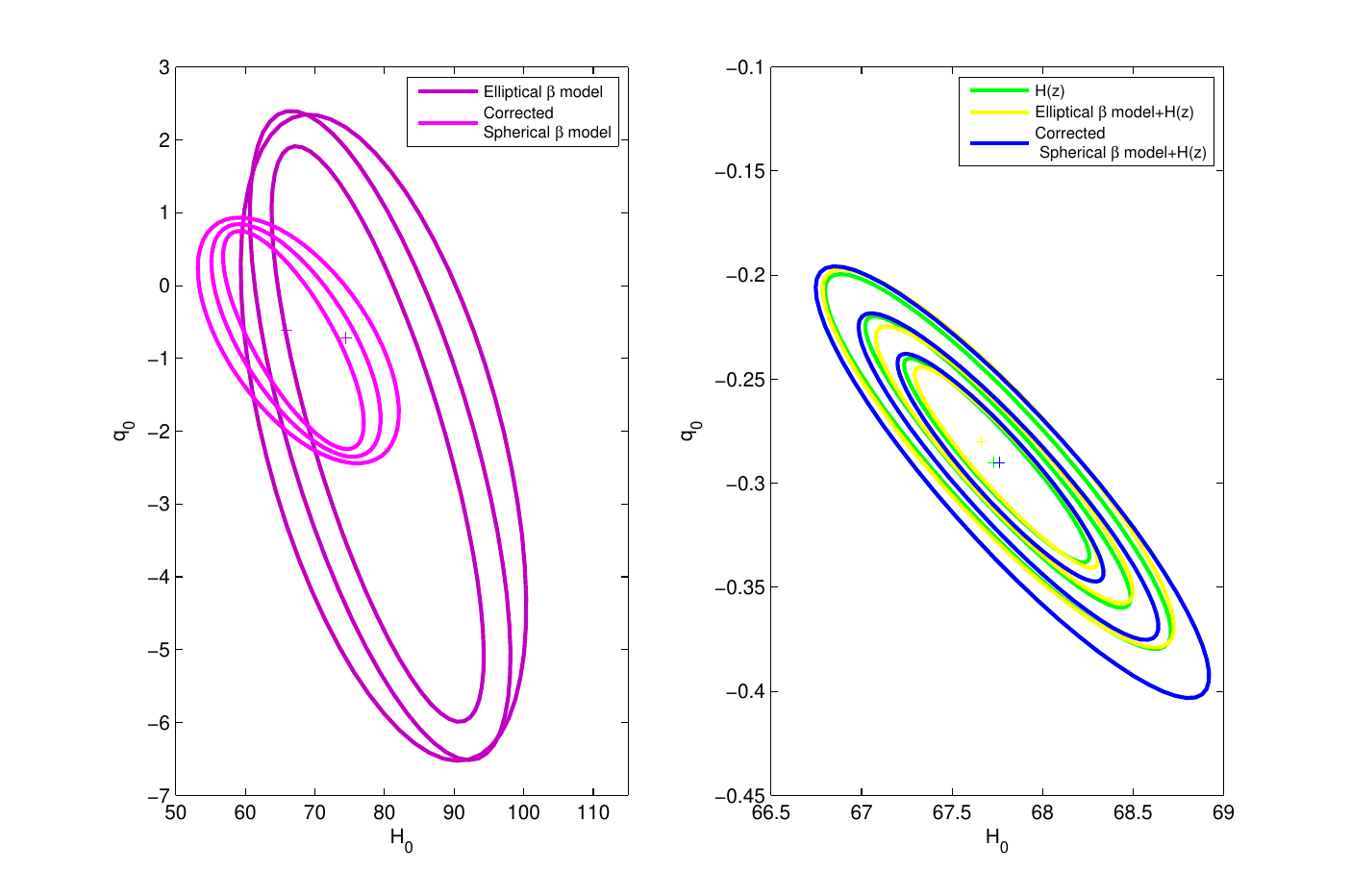}
  \caption{68.3\%, 95.4\%, and 99.7\% confidence region of the ($H_0$, $q_0$ ) plane from $d_A$, $H(z)$, $d_A+H(z)$ observation data combined with the Cosmography \uppercase\expandafter{\romannumeral1}, the + dots in responding color represent the best fitting values for $H_0$, $q_0$.}
\end{figure}

Meanwhile, We combine $d_A$ and $H(z)$ data with Cosmography \uppercase\expandafter{\romannumeral2}, and constraining the parameters. Also getting the parameters $H_0$, $q_0$, $j_0$ best fitting values, statistic mean and error in $68.3\%$($\Delta {\chi ^2} \le 1$) confidence region (see table 2), and probability density graph, confidence region of plane ($H_0$, $j_0$)(see figure 3, 4)

\begin{table*}
\footnotesize
 \centering
   \caption{The best fitting and mean values of the cosmological parameters from $d_A$, $d_c$ and $H(z)$observation data combined with the Cosmography \uppercase\expandafter{\romannumeral1}, in 68.3\% confidence region.}
  \vspace{0.3cm}
  \begin{tabular}{@{}cccccc@{}}
  \hline
  & Data parameters&$H_0$&$q_0$&$j_0$&$\chi _{\min }^2$\\ \hline
   Elliptical-$\beta$ model &Best fitting &78.65   &    -2.21      &    25.57    &    24.11$/$24\\
                            &Mean&79.02$\pm$4.34 & -2.04$\pm$1.13   &   24.55$\pm$24  \\ \hline
   Corrected Spherical &Best fitting &70.44  &-1.65    &    10.99   &    31.56$/$32\\
                            -$\beta$ model&Mean&72.5$\pm$3.1 & -1.99$\pm$0.21   &   16.04$\pm$6.12  \\ \hline
  $H(z)$ &Best fitting &69.48   &    -0.64      &    1.84    &    28.36$/$28\\
                            &Mean&69.13$\pm$0.24 & -0.57$\pm$0.04   &   1.28$\pm$0.33  \\ \hline
   Elliptical-$\beta$ model &Best fitting &69.51   &    -0.64      &    1.86     &    53.68$/$52\\
                           $+H(z)$  &Mean&69.63$\pm$0.25 & -0.66$\pm$0.04   &   2.1$\pm$0.39  \\ \hline
   Corrected Spherical &Best fitting &69.44   &    -0.64      &    1.87     &    63.64$/$60\\
                            -$\beta$ model $+H(z)$&Mean&69.11$\pm$0.26 & -0.57$\pm$0.05   &   1.31$\pm$0.4  \\
  \hline
\end{tabular}
\end{table*}

\begin{figure*}[htbp]
\includegraphics[scale=0.7]{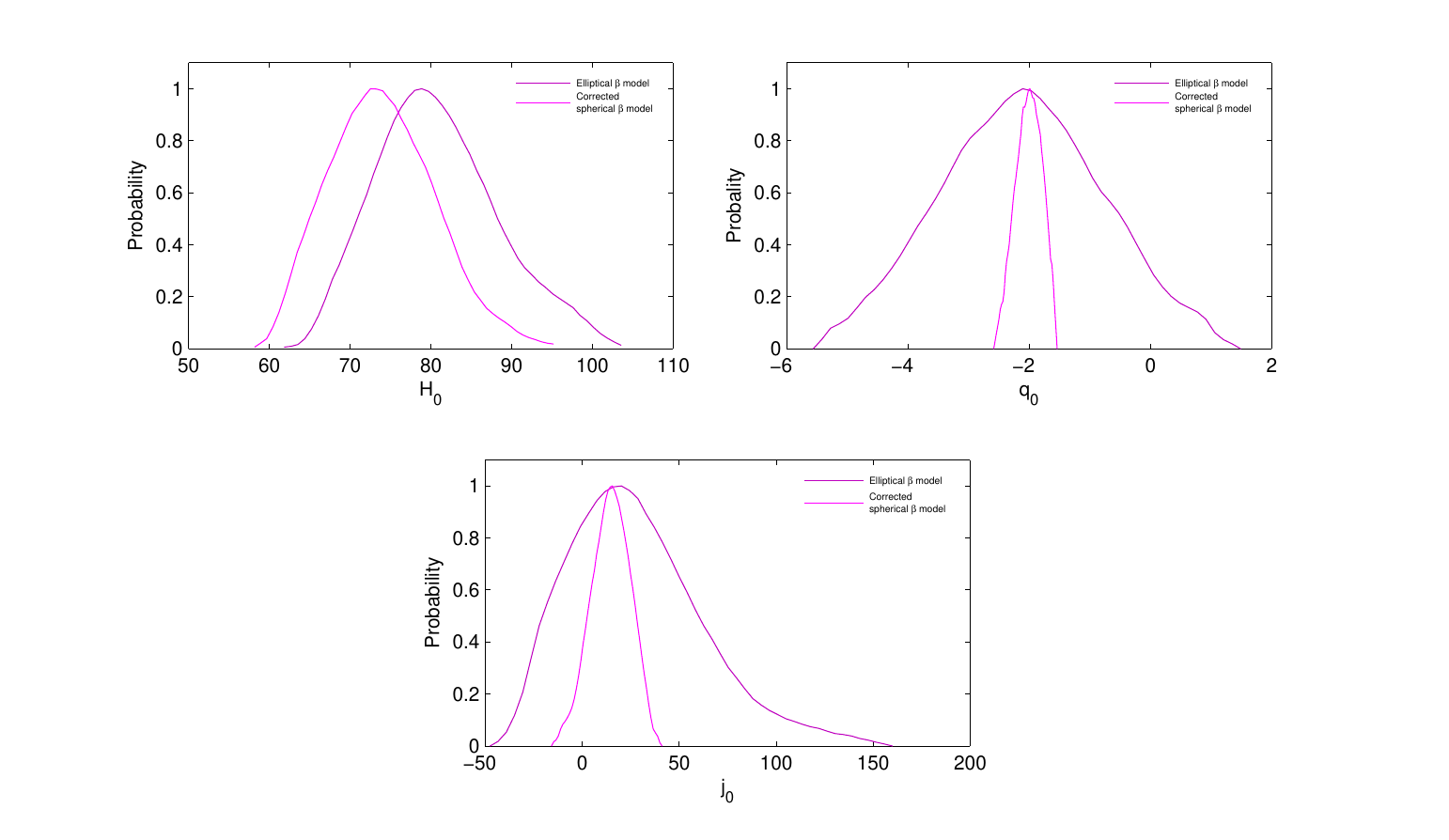}
  \caption{One-dimensional probability distribution graph for the parameter $H_0$, $q_0$ and $j_0$ from  $d_A$ observation data combined with the Cosmography \uppercase\expandafter{\romannumeral2}}.
\end{figure*}

\begin{figure*}[htbp]
\includegraphics[scale=0.7]{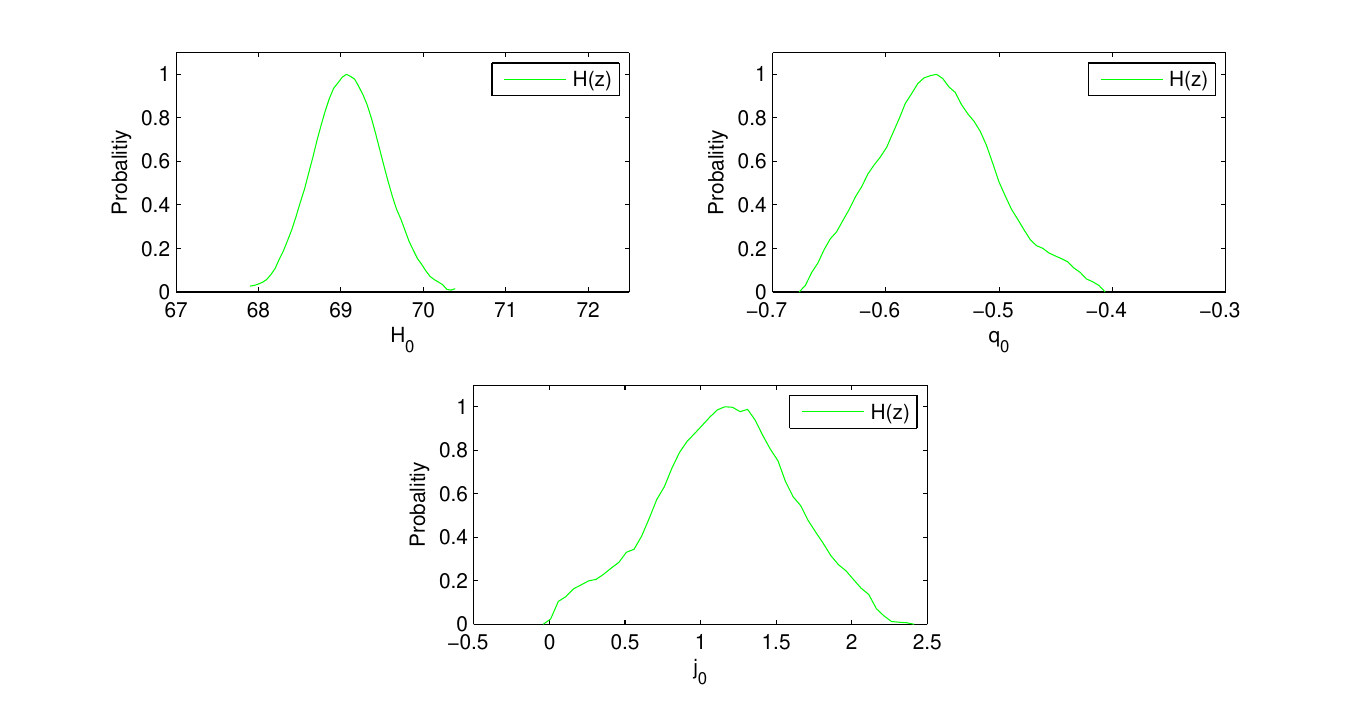}
  \caption{One-dimensional probability distribution graph for the parameter $H_0$, $q_0$ and $j_0$ from  $H(z)$,$d_A+H(z)$ observation data combined with the Cosmography \uppercase\expandafter{\romannumeral2}}.
\end{figure*}

\begin{figure}[tbp]
\centering
\includegraphics[angle=0,scale=0.35]{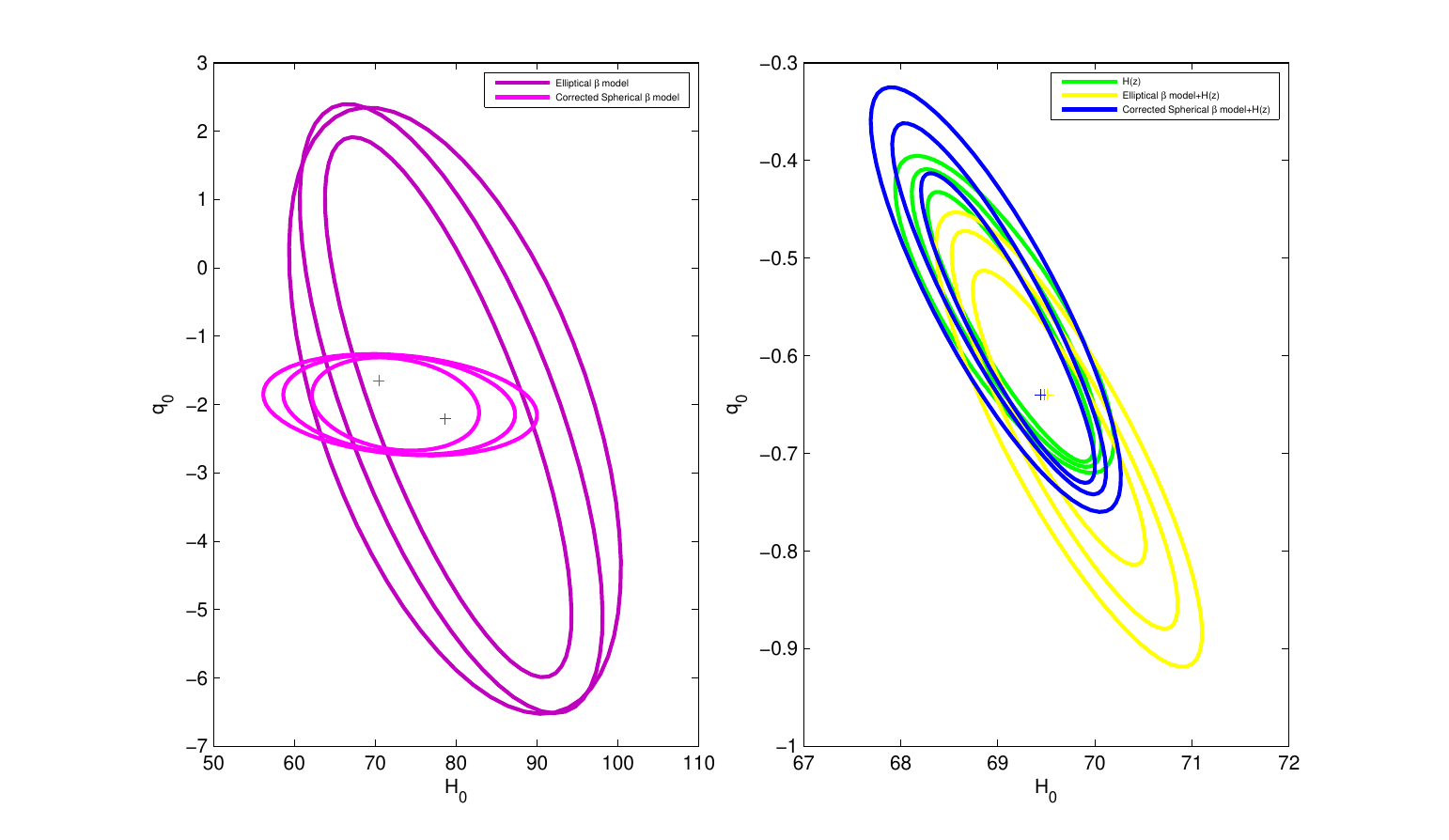}
  \caption{68.3\%, 95.4\%, and 99.7\% confidence region of the ($H_0$, $q_0$ ) plane from $d_A$, $H(z)$, $d_A+H(z)$ observation data combined with the Cosmography \uppercase\expandafter{\romannumeral2}, the + dots in responding color represent the best fitting values for $H_0$, $q_0$.}
\end{figure}

\section{The parameter values analysis and comparison}

From figure 3 (left-hand side), we can see the parameters ($H_0$,$q_0$) best fitting values is (74.43, -0.72), (65.91,-0.61), which is gotten by Elliptical $\beta$ model, spherical $\beta$ model $d_A$ data and Cosmography \uppercase\expandafter{\romannumeral1}, and the range of deceleration factor $q_0$ in 99.7\% confidence region ($\Delta {\chi ^2} \le 4$) is small difference in the results from current SNe Ia data and CMB data combined with $\Lambda CDM$ or $\omega CDM$ \citep{19}; From figure 3 (right-hand side), we can see the parameters ($H_0$,$q_0$) best fitting values is (67.73, -0.29), which is gotten by $d_c$ data from $H(z)$ data and Cosmography \uppercase\expandafter{\romannumeral1}, and the range of deceleration factor $q_0$ in 99.7\% confidence region is large difference in the results from current SNe Ia data and CMB data combined with $\Lambda CDM$ or $\omega CDM$. Because the redshift range is $0.023 \le z \le 0.784$, $0.142 \le z \le 0.826$ for Elliptical $\beta$ model and spherical $\beta$ model $d_A$ data respectively, their redshift all less than 1, but the redshift range is $0.07 \le z \le 1.75$ for $d_c$ data from $H(z)$, so $d_A$ data can more accurately constrain the parameters to use Cosmography \uppercase\expandafter{\romannumeral1}.

As well, From figure 6 (left-hand side), we can see the parameters ($H_0$,$q_0$) best fitting values is (78.65, -2.21), (70.44,-1.65), which is gotten by Elliptical $\beta$ model, spherical $\beta$ model $d_A$ data and Cosmography \uppercase\expandafter{\romannumeral2}, and the range of deceleration factor $q_0$ in 99.7\% confidence region ($\Delta {\chi ^2} \le 4$) is large difference in the results from current SNe Ia data and CMB data combined with $\Lambda CDM$ or $\omega CDM$. From figure 6 (right-hand side), we can see the parameters ($H_0$,$q_0$) best fitting values is (69.48, -0.64), which is gotten by $d_c$ data from $H(z)$ data and Cosmography \uppercase\expandafter{\romannumeral1}, and the range of deceleration factor $q_0$ in 99.7\% confidence region is small difference in the results from current SNe Ia data and CMB data combined with $\Lambda CDM$ or $\omega CDM$. Because the redshift range is $0.07 \le z \le 1.75$ for $d_c$ data from $H(z)$, so it can more accurately constrain the parameters to use Cosmography\uppercase\expandafter{\romannumeral2}.
From $d_c$ data, we get ${H_0} = 69.13 \pm 0.24km \cdot {s^{ - 1}} \cdot Mp{c^{ - 1}}$, correspond to 0.4\% uncertainty.

\section{Conclusion}

We can use cosmography and galaxy clusters $d_A$, Hubble parameters $H(z)$ data to constrain Hubble constant $H_0$, deceleration factor $q_0$, etc. For $H(z)$ data, because it does not need to be calibrated like other observation data, we consider Hubble parameter $H_0=69.13\pm 0.24{\kern 1pt} km{\kern 1pt}  \cdot {s^{ - 1}} \cdot Mp{c^{ - 1}}$, deceleration factor $q_0=-0.57\pm0.04 $, acceleration rate $j_0=1.28\pm0.33$ by $H(z)$ data seems to be more persuasive. we guess $w$ in $\omega CDM$ model may be greater than $-1$ by the value of $j_0$ \citep{32,33}. Of course, we need more $H(z)$ data and other precise observation to confirm further the values of cosmology kinematical state parameters. In future, we expect to obtain more $H(z)$ data, accurate $d_A$ and other data, and then constraining the Hubble constant $H_0$ again, to get more accurate value. Meanwhile, we will use current observation data, such as SNe Ia, galaxy clusters, Hubble parameters $H(z)$, etc. Taking Guassian Process (GP) method to get a series data\citep{34,35}, and then directly constrain cosmological parameters, to get more accurate fitting values.
\section{Acknowledgments}
This project is supported by the High-level Scientific Research Foundation for the
introduction of talent(Grant No. 20160375114 ) and Science and Technology Talents Plan of the department of education of Guizhou province ( KY[2016]090 ) .
\section*{References}
%
%\bibliography{2}

%\input{main.bbl}

\end{document}